\documentclass{article}
\usepackage{color}
\usepackage{spconf,amsmath,graphicx,url,bbm}
\usepackage{amssymb}
\usepackage{tabularx,esdiff,algorithm, mathtools,xparse}
\title{Practical Noise Simulation for RGB Images}
\name{{Saeed Ranjbar Alvar and Ivan V. Baji\'c}}
\address{School of Engineering Science, Simon Fraser University, Burnaby, BC, Canada 
}
\begin{document}
%

\maketitle
\begin{abstract}
This document describes a noise generator that simulates realistic noise found in smartphone cameras. The generator simulates Poissonian-Gaussian noise whose parameters have been estimated on the Smartphone Image Denoising Dataset (SIDD). The generator is available online, and is currently being used in compressed-domain denoising exploration experiments in JPEG AI. 
\end{abstract}
%


\section{Introduction}
\label{sec:intro}

Sensing is a noisy process. 
Digital imaging sensors  record scenes by converting the number of photons that hit individual photosites to digital intensities. Various steps in forming a digital image introduce noise into the process, making the image captured at the sensor noisy.  Noise in the captured image is primarily the result of two factors. The first factor is photon noise, which is caused by the randomness in number of photons arriving at a particular photosite over a time unit. The second factor is related to the sensor noise~\cite{thesis}. Noise modeling and image denoising have been active research topics for years and the recent efforts try to address the practical challenges related to modeling and reducing the noise in the fast growing mobile and camera industries.

\section{Noise Generator}
There are many methods in the literature to model the noise in the images obtained from digital imaging sensors. The most widely studied model is independent and identically distributed (iid, or white) Gaussian noise, which contaminates each pixel with a random variable drawn from a zero-mean Gaussian distribution of a given variance. 
Practical noise is more involved than this simple model: it may be signal-dependent, and its strength (variance) might not be uniform across the image. 

Poissonian-Gaussian model~\cite{Foi} is a popular signal-depen-dent noise model that is used for modeling the noise in raw-RGB images. The noisy observation at pixel $x$ is formed as:
\begin{equation}
    z(x) = y(x) + n(x), 
\end{equation}
where $x\in X$ is the pixel position in the domain $X$, $z: X \longrightarrow \mathbb{R}$ is the observed noisy signal,  $y: X \longrightarrow \mathbb{R}$ is the original clean signal, $n: X \longrightarrow \mathbb{R}$ is the noise. The noise is composed of two components:
\begin{equation}
    n(x) = n_p(y(x)) + n_g(x),
\end{equation}
Where $n_p$ is the Poissonian signal-dependent component and $n_g$ is the Gaussian signal-independent component. These two components are characterized as follows:
\begin{align}
\frac{1}{a} (y(x) +  n_p(y(x)))  &\sim \mathcal{P}\left(\frac{y(x)}{a}\right), \label{eq:noise_model_a}\\
n_g(x) &\sim \mathcal{N}(0,b), 
\label{eq:noise_model_b}
\end{align}
where $a>0$ and $b\geq0$ are the parameters of the Poisson distribution ($\mathcal{P}$) and the Gaussian distribution ($\mathcal{N}$), respectively. It can be shown that the variance of $z(x)$ has the form:
\begin{equation}
    \mathrm{var}(z(x)) = a\cdot y(x) + b,
    \label{eq:noise_variance}
\end{equation}
where $\mathrm{var}(\cdot)$ computes the variance of the signal. Authors in~\cite{Foi} also proposed a method to estimate $a$ and $b$ from single noisy raw-RGB image.  

A raw-RGB image represents a sensor image whose RGB values correspond to the color sensitivity of the sensor. The datasets used for training deep learning models contain standard RGB (sRGB) images in general. Therefore, raw-RGB images obtained from the sensors are converted to sRGB images by applying signal processing techniques before being used in practice. 
When raw-RGB images are transformed into sRGB colour space, the distribution of noise in raw-RGB images changes significantly. Hence, the estimated noise parameters for raw-RGB images do not hold for sRGB images. In this work, we develop a method to estimate the parameters of the Poissonian-Gaussian noise model for sRGB images in Smartphone Image Denoising Dataset (SIDD)~\cite{SIDD}, and use the distribution of the estimated parameters to obtain a practical noise generator. The obtained noise generator can be used for generating realistic noise samples. In addition, the obtained noise generator can be used to add multiple realizations of the modeled noise to any clean image, which can create a large dataset for training deep learning models.

The color components in sRGB images are derived from raw-RGB images using different processes. Hence, the noises in color components are different although they are captured with the same sensor, and the parameters of the noise are estimated for each color component separately. To estimate the parameters of the noise for sRGB images in SIDD, we used the estimator from~\cite{Foi} whose implementation is provided by the authors.\footnote{\url{https://webpages.tuni.fi/foi/sensornoise.html#ref_software}} The estimator is applied to the 500$\times$500 patches cropped from high resolution images in each of 160 scenes in the SIDD dataset. 

\begin{figure}[t]
\centering
\includegraphics[width=0.4\textwidth]{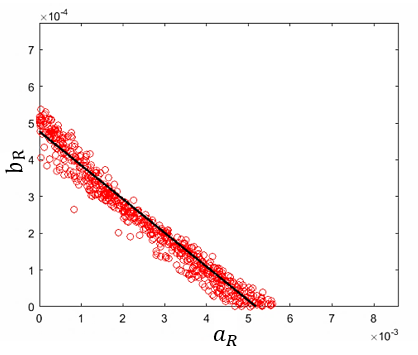}
\caption{A fitted line example for Red components of the images in one of the scenes in the SIDD dataset.}
\label{fig:line}
\end{figure}

For some scenes, the estimator returns $a$ and/or $b$ parameters that are less than zero. This contradicts the model assumed  in~(\ref{eq:noise_model_a})-(\ref{eq:noise_model_b}). To get around this problem, we proceeded as follows. We fit a straight line to the estimated samples, defined as:
\begin{equation}
    b = m\cdot a + c,    
\end{equation}
where $m$ and $c$ are the slope and the intercept of the fitted line. The fitted line captures the relation between $a$ and $b$ as shown in the example in Fig.~\ref{fig:line}. The histogram of the slopes for different scenes is recorded and later used in the noise generator.
The intercept $c$ corresponds to the variance of a Gaussian noise when there is no Poissonian component present. 
Hence, we recorded the histogram of noise variances, computed from the differences of noisy and clean images, and used it subsequently to sample intercepts $c$. One such histogram was recorded for each color component. Besides that, we also recorded histograms of positive $a$ estimates for each color component. 

\begin{figure*}[t]
\centering
\includegraphics[width=5.5cm]{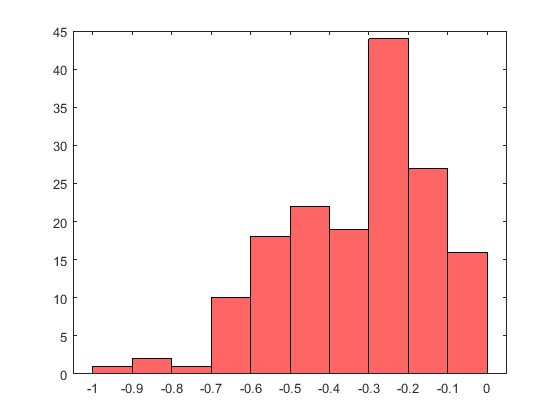}
\includegraphics[width=5.5cm]{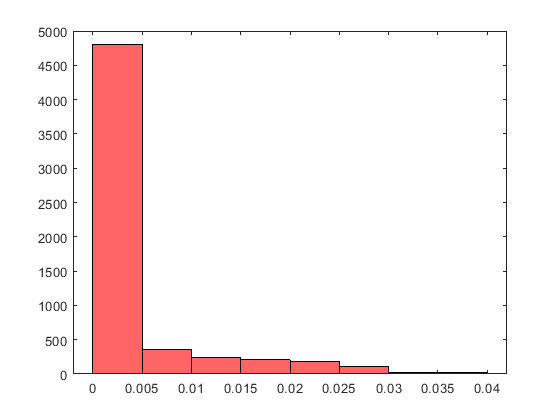}
\includegraphics[width=5.5cm]{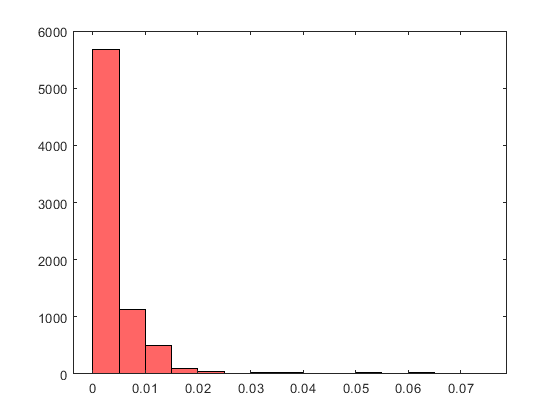} \\
\includegraphics[width=5.5cm]{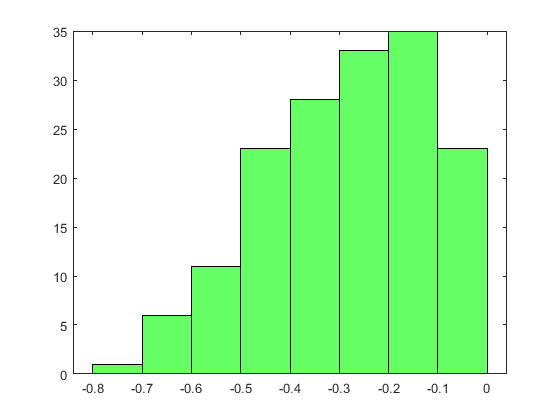}
\includegraphics[width=5.5cm]{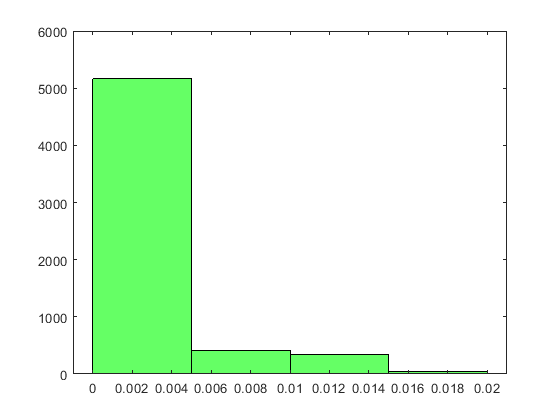}
\includegraphics[width=5.5cm]{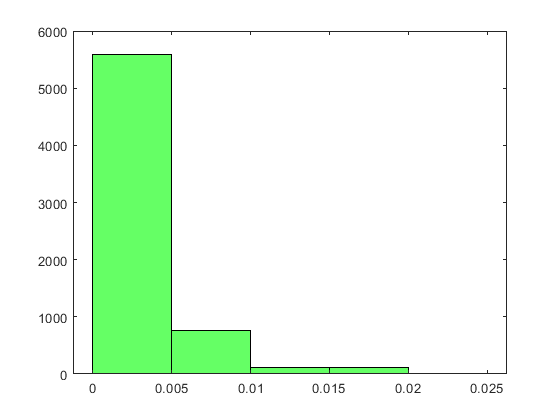} \\
\includegraphics[width=5.5cm]{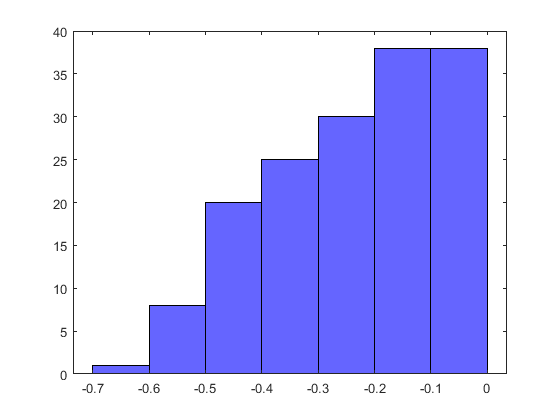}
\includegraphics[width=5.5cm]{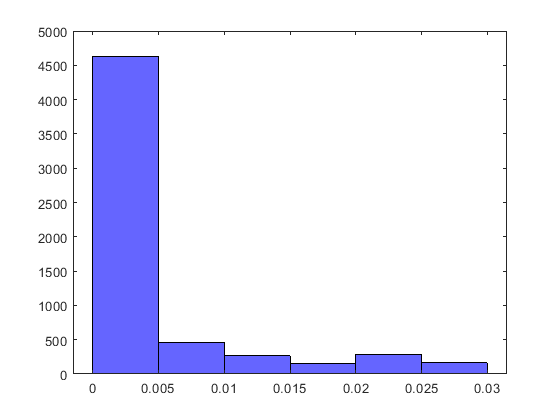}
\includegraphics[width=5.5cm]{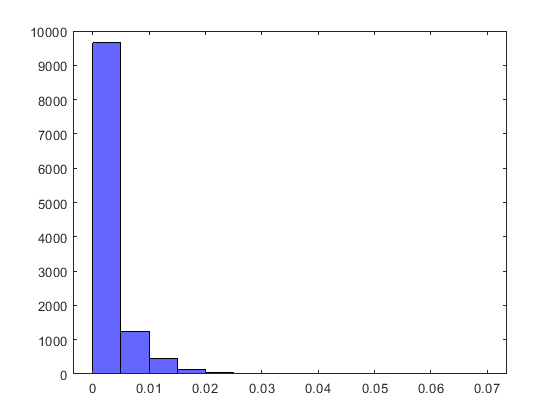} \\
\caption{Histograms of the estimated parameters.Top row: Red, Middle row: Green, Bottom row: Blue. Left column: histogram of slopes, Middle column: histogram of intercepts, Right column: histogram of $a>0$}
\label{fig:histograms}
\end{figure*}

If we draw a sample from the histogram of slopes and draw a sample from the histogram of noise variances, we have the slope ($m'$) and the intercept ($c')$ of the line $b= m'\cdot a + c'$. To obtain a specific point $(a',b')$ on this line, we draw a sample $a'$ from the histogram of positive $a$ estimates, 
then compute $b'=m'\cdot a'+c'$. If $b'$ happens to be negative, $m'$, $c'$ and $a'$ are resampled until a positive $b'$ is obtained. 
The histograms of the slope, intercept and positive $a$ estimates for different color components are shown in Fig.~\ref{fig:histograms}. 

\begin{figure*}[t]
\centering
\includegraphics[width=0.8\textwidth]{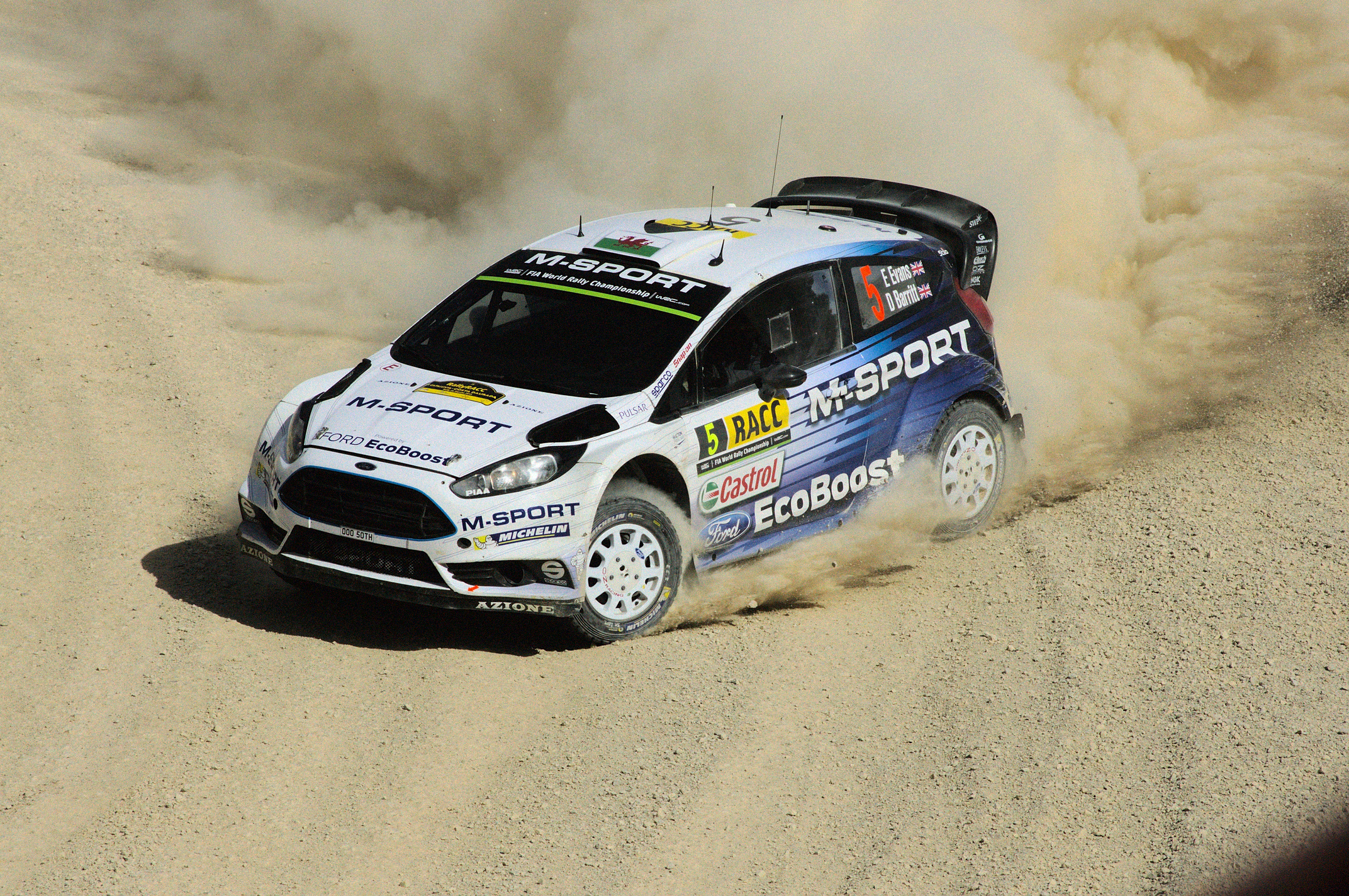} \\
\vspace{10pt}
\includegraphics[width=0.8\textwidth]{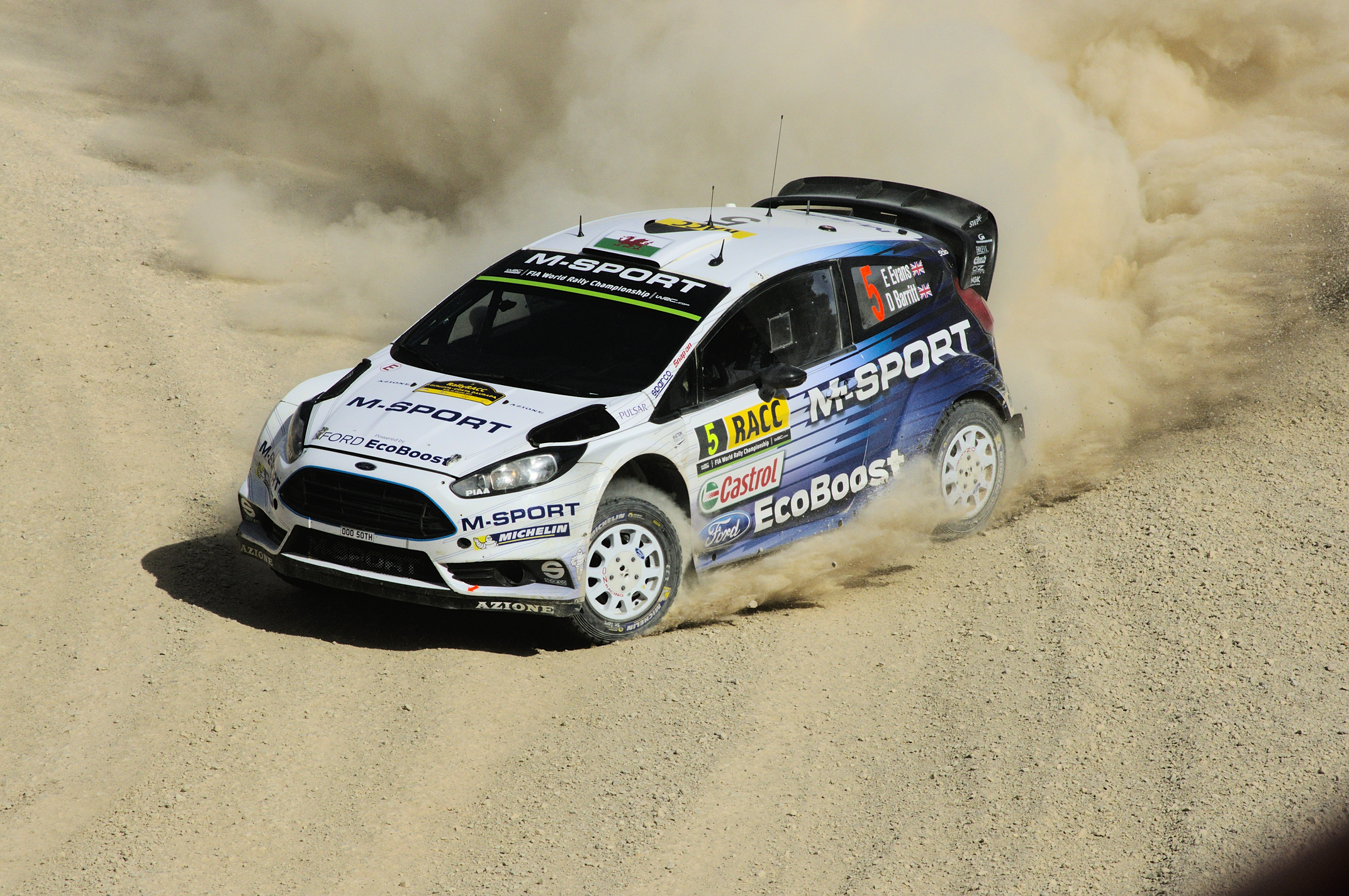}

\caption{An example of a noisy image with  $\mathbf{a} = [0.0002, 0.0001, 0.0001]$, $\mathbf{b}=[0.0030, 0.0004, 0.0009]$. Top: Noisy image, Bottom: Clean image}
\label{fig:noise_example}
\end{figure*}

In the proposed noise generator, noise parameters for (R,G,B) color components are obtained using the histograms shown in Fig.~\ref{fig:histograms}. The shown histograms represent the distribution of the corresponding parameters. Inverse transform sampling~\cite{inverse_sampling} is used to sample the mentioned histograms. By sampling the histograms and computing the $(a,b)$ pairs for each color component, we obtain $\mathbf{a} = (a_R,a_G,a_B)$ and $\mathbf{b} = (b_R,b_G,b_B)$, where $a_R$, $a_G$, $a_B$ are the estimated $a$ parameters for red, green and blue color channels, respectively, and $b_R$, $b_G$, $b_B$ are the estimated $b$ parameters for red, green and blue color channels, respectively. An example of the generated noisy image using the noise generator for the given clean image is shown Fig.~\ref{fig:noise_example}.

\section{Usage}
The noise generator is available online.\footnote{ \url{https://github.com/SFU-Multimedia-Lab/noise_generator}} 
Multiple noisy images ($N$) can be obtained from a given clean image in \texttt{IMG\_DIR}  by running the code as:

\texttt{python ./noise\_sampling.py 
--{img\_dir} \\{\$IMG\_DIR} 
--{n\_obs} \$N --{out\_dir} {\$OUT\_DIR}
}

The above code generates $N$ noisy images for the given image (\texttt{IMG\_DIR}) using the provided noise generator and saves them in (\texttt{OUT\_DIR}). The proposed noise generator is part of the JPEG AI common training and test conditions~\cite{JPEG-AI_CTTC} for the compressed-domain denoising task. 

 


\bibliographystyle{IEEEbib}
\bibliography{refs}

\end{document}